\documentclass{ecai} 

\usepackage{latexsym}
\usepackage{amssymb}
\usepackage{amsmath}
\usepackage{amsthm}
\usepackage{booktabs}
\usepackage{enumitem}
\usepackage{graphicx}
\usepackage{color}
\usepackage{tikz}
\usepackage{bm}
\usepackage{amsmath}
\usepackage[shortcuts]{extdash}
\usetikzlibrary{arrows.meta, positioning, shapes.geometric, calc, fit, backgrounds, automata}
\usepackage{dblfloatfix}

\newcommand{\BibTeX}{B\kern-.05em{\sc i\kern-.025em b}\kern-.08em\TeX}

\begin{document}


\begin{frontmatter}


\paperid{} 


\title{Preliminary Investigation into Uncertainty-Aware Attack Stage Classification}


\author[A]{\fnms{Alessandro}~\snm{Gaudenzi}}

\author[A]{\fnms{Lorenzo}~\snm{Nodari}}

\author[B]{\fnms{Lance}~\snm{Kaplan}}

\author[C]{\fnms{Alessandra}~\snm{Russo}}

\author[D]{\fnms{Murat}~\snm{Sensoy}}

\author[A]{\fnms{Federico}~\snm{Cerutti}}

\address[A]{University of Brescia, Italy}
\address[B]{US Army DEVCOM Army Research Lab, USA}
\address[C]{Imperial College London, UK}
\address[D]{Amazon Alexa AI, UK}


\begin{abstract}
Advanced Persistent Threats (APTs) represent a significant challenge in cybersecurity due to their prolonged, multi-stage nature and the sophistication of their operators. Traditional detection systems typically focus on identifying malicious activity in binary terms \--- benign or malicious \--- without accounting for the progression of an attack. However, effective response strategies depend on accurate inference of the attack’s current stage, as countermeasures must be tailored to whether an adversary is in the early reconnaissance phase or actively conducting exploitation or exfiltration.
This work addresses the problem of attack stage inference under uncertainty, with a focus on robustness to out-of-distribution (OOD) inputs. We propose a classification approach based on Evidential Deep Learning (EDL), which models predictive uncertainty by outputting parameters of a Dirichlet distribution over possible stages. This allows the system not only to predict the most likely stage of an attack but also to indicate when it is uncertain or the input lies outside the training distribution.
Preliminary experiments in a simulated environment demonstrate that the proposed model can accurately infer the stage of an attack with calibrated confidence while effectively detecting OOD inputs, which may indicate changes in the attackers' tactics. These results support the feasibility of deploying uncertainty-aware models for staged threat detection in dynamic and adversarial environments.
\end{abstract}

\end{frontmatter}


\section{Introduction}
The contemporary cyber threat landscape is defined by a rapidly expanding attack surface, the integration of artificial intelligence into offensive operations, and the increasing frequency of targeted intrusions. Among these, Advanced Persistent Threats (APTs) pose a significant challenge~\cite{alshamrani2019survey,wang2024combating}. APTs are characterised by their extended duration, careful orchestration, and emphasis on stealth and persistence. They are often executed by highly capable adversaries with the objective of maintaining long-term access, exfiltrating sensitive data, or positioning themselves for future operations.

A central challenge in responding to such threats is the inference of the attack stage ~\cite{wilkens2021multi}. Unlike binary classification of traffic as benign or malicious, identifying the phase of an ongoing intrusion is both more granular and operationally significant. Defensive strategies vary considerably depending on whether an adversary is conducting reconnaissance, delivering a payload, exploiting a vulnerability, or already operating within the target environment. Inaccurate stage inference can lead to suboptimal mitigation: premature intervention may trigger evasion, whereas delayed response increases impact.

The Lockheed-Martin cyber kill chain~\cite{hutchins2011intelligence} provides a canonical model of staged attack progression. It defines phases such as reconnaissance, weaponisation, delivery, exploitation, installation, and command-and-control. The task of mapping observed system or network activity to these abstract stages is non-trivial, particularly under operational constraints and adversarial adaptation.

A key complication in this classification task arises from the presence of out-of-distribution (OOD) inputs ~\cite{talpini2024enhancing}. These include previously unseen tactics, novel malware behaviours, or benign anomalies that fall outside the support of the training data. Classifiers that lack mechanisms to detect and account for such inputs are prone to overconfidence and misclassification. This is particularly problematic in high-stakes contexts like APT response, where incorrect inferences regarding the stage of an attack can lead to cascading failures in containment and remediation.

Consequently, models deployed for attack stage inference must not only be accurate under familiar conditions but also uncertainty-aware. That is, they should indicate when a prediction is unreliable or when an input lies outside the model's training distribution. Incorporating uncertainty into predictions allows decision-makers to calibrate trust in model outputs and to defer or adapt responses in the presence of ambiguity.

This work expands upon \cite{DBLP:conf/itasec/GaudenziNVGDR025} and addresses the problem of attack stage inference under uncertainty. The classifier ingests a feature vectors that combines high-level system-state indicators with auxiliary labels, formulates stage classification as a probabilistic task over a structured threat progression model and leverages predictive distributions to quantify confidence. Particular attention is given to handling OOD inputs through mechanisms that explicitly model epistemic uncertainty, thereby enabling the system to recognise when the input does not conform to the known stages of adversarial behaviour. To this end, we employ Evidential Deep Learning (EDL), which allows the model to represent both the predicted class distribution and the associated uncertainty via a Dirichlet distribution. This approach facilitates a principled means of detecting anomalous or unfamiliar inputs without requiring an explicit OOD dataset during training.

Preliminary results obtained from a controlled simulation environment demonstrate the feasibility of inferring the stage of an attack using this method. The model is able to distinguish between different phases of the attack lifecycle with calibrated confidence scores and, crucially, maintains robustness in the presence of OOD inputs. These initial findings suggest that uncertainty-aware classification using EDL provides a viable foundation for real-world systems that must operate reliably in the face of evolving and incomplete threat intelligence.

This paper is organized as follows: Section 2 presents the background and related work relevant to our study, including the framework used, the attack modeling background, reward machines and evidential deep learning explanation. Section 3 describes the methodology and experimental setup used in our research, as the simulated environment, data collection and model architecture. Section 4 presents and analyzes the experimental results. Finally, Section 5 concludes the paper and discusses future research directions.

\section{Background}
\subsection{Microsoft CyberBattleSim}

Microsoft CyberBattleSim\footnote{\url{https://github.com/microsoft/CyberBattleSim} (on 15 Jul 2025).} represents a significant advancement in cybersecurity research by providing a simulated network environment designed for experimentation and training. Built on OpenAI's Gymnasium framework, CyberBattleSim abstracts the complexities of real networks by representing machines as nodes in a graph structure. This abstraction makes the platform accessible for researchers and developers who do not need a fully implemented network infrastructure to use it. Reinforcement learning agents act as attackers attempting to compromise the network through various actions. These include local attacks that can expose passwords and reveal sensitive data like credentials, as well as remote vulnerabilities that allow direct node compromise. Once a node is compromised, agents can move laterally using protocols such as SSH or RDP \cite{norris2025enhancing}.
The platform has proven valuable for cyber defense exercises and capture-the-flag scenarios, allowing practitioners to explore different types of exploits on system vulnerabilities. This works particularly well because reinforcement learning (RL) algorithms excel at cybersecurity tasks due to their exploration capabilities, which help discover previously unknown attack and defense scenarios \cite{10068930}. 
CyberBattleSim's key advantage lies in its ability to function as a \textit{physics engine} for the cyber domain, providing dynamic feedback that static datasets cannot offer. This makes it particularly valuable for addressing the critical need for immediate decision-making in modern cyberattacks and for developing incident response skills \cite{andrewspillard2022}.

\subsection{Attack models}
Attack models represent formal abstractions of adversary strategies and behaviors. They capture the goals an attacker seeks to achieve, the decision points encountered along the way, and the sequence of tactics and techniques that may be employed. By defining these patterns in a structured form, attack models enable security teams to anticipate potential threat paths, simulate attack scenarios, and validate the effectiveness of defensive controls.

The MITRE ATT\&CK Flow\footnote{\url{https://center-for-threat-informed-defense.github.io/attack-flow} (on 15 Jul 2025).} is a structured framework designed to visualise systematically and model adversary behaviours and attack sequences. It allows security teams to document, analyse, and communicate the progression of attacks using a flowchart-like representation, connecting individual tactics, techniques, and procedures (TTPs) from the MITRE ATT\&CK knowledge base. By mapping these sequences, organisations can better understand how an adversary moves through an attack lifecycle, identify potential defence gaps, and improve detection and response strategies. The framework enhances situational awareness, enabling more robust security postures and collaborative threat analysis.

\subsection{Attack Stage Inference via Reward Machine State Estimation}

The problem of attack stage classification can be framed as the task of inferring the abstract progression of an adversary through a series of behavioural states based on a sequence of observed actions. This is analogous to identifying the underlying reward machine state of a RL agent operating within an environment, such as an attacker trying to achieve a specific objective. 

A \emph{reward machine} is a finite-state automaton used to specify structured reward functions for RL agents \cite{icarte2018using}. It consists of:
\begin{itemize}
  \item \textbf{States:} abstract representations capturing the agent's progression through a task.
  \item \textbf{State transition function:} rules for moving between states based on logical conditions derived from observations.
  \item \textbf{Labelling function:} a mapping from low-level environmental features to symbolic propositions used in transition conditions.
\end{itemize}

Reward machines increase the expressiveness of task specifications and improve the interpretability of agent behaviour. They also facilitate the reuse of learned behaviours across structurally similar tasks. From a cybersecurity standpoint, this abstraction enables the modelling of attacker behaviour in terms of symbolic subgoals, akin to stages in the cyber kill chain, while omitting low-level artefacts such as system logs or process-level telemetry (\textbf{observations}). The labelling function then abstracts such low-level artefacts into symbolic propositions (\textbf{labels}) which might indicate that a new stage of the attack has begun, e.g., in the Uber breach, when the attackers compromised the credentials of an external contractor (abstract label), they were then able to proceed with lateral movements (new stage of the attack).\footnote{\url{https://center-for-threat-informed-defense.github.io/attack-flow/ui/?src=..\%2fcorpus\%2fUber\%20Breach.afb} (on 11 June 2025).}

This analogy is particularly effective in honeypot environments. Unlike production systems where benign activity dominates and malicious activity is sparse, honeypots are designed to attract and isolate unauthorised interactions. As a result, the sequence of actions observed in a honeypot is typically devoid of "sane" or legitimate behaviour, consisting instead of purely adversarial traces. This simplifies the modelling problem by reducing the behavioural variance and ensuring that all recorded activity is relevant for stage inference. The attacker’s actions can thus be treated as the trace of an RL agent operating in a known environment, where the reward machine states correspond to phases in the adversary’s operational logic.

\subsection{Evidential Deep Learning}
Inferring the correct reward-machine state from behavioural traces \--- which includes both observations and labels \--- remains subject to ambiguity, particularly in the presence of incomplete or atypical sequences. This necessitates the use of classification methods that are not only accurate but also capable of representing uncertainty. 
We summarise an approach based on Dirichlet distributions to model classification uncertainty, following \cite{EDLGEN}. Rather than relying on softmax outputs as point estimates, this method predicts parameters of a Dirichlet distribution over class probabilities, allowing uncertainty to be quantified alongside predictions.

The Dirichlet distribution $D(\mathbf{p}|\bm{\alpha})$ over the class probability vector $\mathbf{p}$ is defined by parameters $\bm{\alpha} = [\alpha_1, \ldots, \alpha_K]$ and has the form
\begin{equation}
D(\mathbf{p}|\bm{\alpha}) = \frac{1}{B(\bm{\alpha})} \prod_{i=1}^K p_i^{\alpha_i - 1},
\end{equation}
for $\mathbf{p}$ in the $K$-dimensional simplex $\mathcal{S}_K$, where $B(\bm{\alpha})$ is the multinomial beta function. The mean and variance of $p_k$ are given by
\begin{equation}
\label{e:pignistic}
\hat{p}_k = \frac{\alpha_k}{S}, \quad Var(p_k) = \frac{\alpha_k (S - \alpha_k)}{S^2 (S + 1)},
\end{equation}
with $S = \sum_{i=1}^K \alpha_i$. A uniform Dirichlet distribution, $D(\mathbf{p}|\bm{1})$, reflects maximum uncertainty.

In this framework, the Dirichlet parameters are interpreted as pseudocounts: $\alpha_k = 1 + e_k$, where $e_k$ represents class-specific evidence inferred from the input. Thus, $S - K$ denotes the total evidence beyond the prior. The mean $\hat{\mathbf{p}}$ is used for classification, while entropy of $\hat{\mathbf{p}}$ quantifies uncertainty.

To estimate $\bm{e}$, the method in \cite{EDLGEN} draws on noise-contrastive estimation and implicit density modelling. For each class $k$, the ratio between the in-class distribution $P_k(\bm{x})$ and a shared out-of-distribution reference $P_{out}(\bm{x})$ is used:
\begin{equation}
\label{eq:log_ratio}
\frac{P_k(\bm{x})}{P_{out}(\bm{x})} = \frac{p(y=k|\bm{x})}{p(y=out|\bm{x})} \left(\frac{1 - \pi_k}{\pi_k} \right),
\end{equation}
where $\pi_k = p(y = k)$ is assumed uniform across classes.

A neural network $\bm{f}(\bm{x}|\theta)$ is trained to approximate the log density ratio $\log(P_k(\bm{x}) / P_{out}(\bm{x}))$ for each class $k$ using the binary Bernoulli loss
\begin{equation}
\label{eq:L1}
\begin{aligned}
\mathcal{L}_1(\theta) = -\sum_{k=1}^K \Big[ &\; \mathbb{E}_{P_k(\bm{x})} \log \sigma(f_k(\bm{x}|\theta)) \\
&+ \mathbb{E}_{P_{out}(\bm{x})} \log(1 - \sigma(f_k(\bm{x}|\theta))) \Big]
\end{aligned}
\end{equation}

Out-of-distribution samples are generated by perturbing training data via a generative adversarial network. These samples differ from training data in input space but remain structurally similar in a learned latent space. After training, $e_k = \exp(f_k(\bm{x}|\theta))$ is treated as evidence for class $k$, and the Dirichlet parameters are computed as $\bm{\alpha} = \bm{e} + \bm{1}$. 

For OOD inputs or outliers, the network produces near-zero evidence and the resulting Dirichlet approaches the uniform prior. For typical in-distribution and correctly classified examples, evidence concentrates on the correct class, such that $e_k > e_j$ for all $j \ne k$.

The vector $\bm{\alpha}$ defines a Dirichlet distribution $D(\bm{p} \mid \bm{\alpha})$ over categorical distributions for the classes of $\bm{x}$. Only one class is correct; if $k$ is the true class, then the marginal distribution over $p_k$ follows a Beta distribution with parameters $\langle \alpha_k, \sum_{j \neq k} \alpha_j \rangle$.

Let $\bm{p}_{-k}$ denote the probabilities for $j \neq k$. The conditional distribution over normalised misclassification probabilities is $\bm{p}'_{-k} \mid p_k \sim D(\bm{p}'_{-k} \mid \bm{\alpha}_{-k})$, where $\bm{p}'_{-k}$ is defined by rescaling $\bm{p}_{-k}$ with $(1 - p_k)$.

To promote uncertainty in misclassifications, a regularisation term is introduced by minimising the Kullback–Leibler divergence between $D(\bm{p}_{-k} \mid \bm{\alpha}_{-k})$ and a uniform Dirichlet:

\begin{equation}
\label{eq:reg_loss}
\mathcal{L}_2(\theta \mid \bm{x}) = \beta \, \mathbb{KL} \left[ D(\bm{p}_{-k} \mid \bm{\alpha}_{-k}) \,\|\, D(\bm{p}_{-k} \mid \bm{1}) \right],
\end{equation}
where $\beta$ controls the strength of the regularisation.

To extract a measure of uncertainty from the Dirichlet distribution in this framework, one may employ the subjective logic approach based on Dirichlet strength \cite{josang2016subjective}. The total strength of the Dirichlet distribution is given by \( S = \sum_{i=1}^K \alpha_i \), representing the sum of pseudocounts. Subjective logic interprets uncertainty as the inverse of this strength, that is, \( u = \frac{K}{S} \), where \( K \) is the number of classes. This formulation reflects that greater total evidence (i.e., higher \( S \)) implies greater confidence in the prediction, and conversely, lower evidence yields higher uncertainty. For instance, in the case of a uniform Dirichlet distribution with \( \alpha_k = 1 \) for all \( k \), the strength is minimal at \( S = K \), and the uncertainty attains its maximum value, \( u = 1 \). This approach allows uncertainty to be quantified explicitly without relying directly on entropy, and is particularly useful in applications where one wishes to separate epistemic uncertainty from aleatoric noise.

\section{Methodology}
\subsection{Switched LAN CTF}
We simulated a switched LAN network, where an attacker begins at a designated entry point and must gain access to a specific target device. The devices are fully interconnected, allowing the attacker to move freely executing lateral movement between any devices in the network.

To reach the target device, which represents a critical target such as a server or database, the attacker must obtain an access credential. Both goal and credential nodes are randomly selected at every simulation run.

The target device is protected by an intrusion prevention system that will detect and block any attempt to access it without the correct credential, resulting in failure. The attacker must navigate the network to find a valid credential before trying to access the target device.

\begin{figure}[tb]
    \centering
    \includegraphics[width=\linewidth]{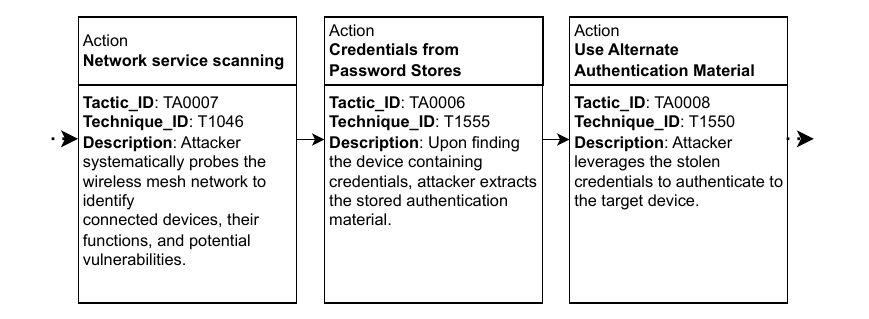}
    \caption{Description in MITRE ATT\&CK Flow of the Switched LAN environment.}
    \label{attack_flow}
\end{figure}
\begin{figure}[tb]
\centering
\begin{tikzpicture}[shorten >=1pt, node distance=3cm, on grid, auto]
  \node[state, initial] (q0) {0};
  \node[state] (q1) [right=of q0] {1};
  \node[state, accepting] (q2) [right=of q1] {2};
  \path[->]
    (q0) edge node {\textbf{c}} (q1)
    (q1) edge node {\textbf{g}} (q2);
\end{tikzpicture}

\caption{Switched LAN CTF Reward Machine, with two labels: \textbf{c}redentials acquired, and \textbf{g}oal achieved.}\
\label{ctf_rm}
\end{figure}
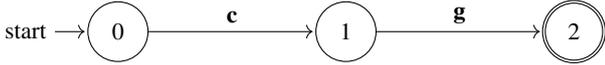

The attack is modeled using MITRE ATT\&CK Flow, which outlines the sequence of actions the attacker may take and the stages of the attack, as shown in figure \ref{attack_flow}. From this modeling is derived a reward machine, presented in figure  \ref{ctf_rm}, with two \textbf{labels}: once the credentials are acquired (\textbf{c}), the attacker needs to find the goal node to exploit (\textbf{g}) to successfully conclude its task.

We selected this switched LAN CTF as experiment environment because it captures essential elements of real-world attacks (credential theft, lateral movement, target exploitation) while maintaining sufficient simplicity for a preliminary investigation setting.

\subsection{Dataset collection}
The experimental dataset was originally introduced in \cite{DBLP:conf/itasec/GaudenziNVGDR025} and was created using a simulated environment based on CyberBattleSim. 
The experimental input data consists of two main components: agent observations of the world state and concatenated labelling function results.

The observation space comprises the union of all valid preconditions for every possible vulnerability present in the network. This state representation focuses on exploitation conditions provided by the framework rather than simple positional information, resulting in a substantially larger observation space compared to basic representations. Action data is excluded from the input since each state transition can only be caused by a single specific action, making the dynamics of state changes contain the necessary information about actions performed and rendering explicit action inclusion redundant.

The labeling function output is included in the dataset for two primary reasons. First, it provides additional context to the model, enhancing its understanding of the environment. Second, it enables us to simulate noise in the labeling function output, which is important for testing the robustness of the approach under realistic conditions where labeling functions may produce imperfect results.

\subsection{EDL-based Model Architecture}


We implemented a neural network that receives input in the form of fixed-length temporal rolling windows extracted from sequential datasets. These windows are structured as 2D tensors where one dimension corresponds to the time sequence length and the other represents the number of features. This windowing strategy enables the model to identify temporal patterns while simultaneously learning feature correlations. The selection of window length depends on the temporal characteristics inherent in the dataset, where extended windows facilitate the capture of long-range dependencies but introduce additional computational overhead.

The neural network architecture used features a hybrid CNN-MLP design structured in three distinct processing phases to enable hierarchical feature extraction and classification. The initial phase applies 2D convolutional operations with configurable filter sizes and channel depths to capture spatial patterns in the input data, followed by max pooling operations that reduce spatial dimensions while preserving salient features. The intermediate phase performs additional convolutional and pooling operations with adjustable parameters to further refine feature representations at multiple scales. The final phase transitions to a fully connected multilayer perceptron consisting of three progressively smaller dense layers. The architecture incorporates flexible hyperparameters including convolutional kernel sizes, pooling window dimensions, stride values, and dense layer neuron counts, allowing for adaptive scaling based on dataset complexity and computational constraints. This hybrid approach leverages the spatial feature extraction capabilities of convolutional layers combined with the representational power of dense layers to achieve robust classification performance across diverse input modalities.


The expectations in Eq.~\ref{eq:L1} are computed by Monte Carlo integration using an equal number of samples from $P_k$ and $P_{out}$. Different from \cite{EDLGEN} \--- that relies on a GAN-based approach with a double discriminator, unsuitable in the present case \--- we use samples from $P_{out}$, which are created using a noise function that stochastically perturbed real input by flipping bits with a specific noise probability of $0.4$. This noise application allowed the model to learn from both unaltered real data and its noisy counterparts. 


We also included two hyperparameters, $w_{\text{real}}$ and $w_{\text{noisy}}$, with the constraint $w_{\text{real}} + w_{\text{noisy}} = 1$ to balance out the two components of the loss function in \eqref{eq:L1}, and we interpreted the $\beta$ hyperparameter of \eqref{eq:reg_loss} as an annealing coefficient.





Such hyperparameters were optimised using Optuna~\cite{2019optuna}, targeting minimal uncertainty over clean validation data. The best-performing configuration for the task considered here was: $w_{\text{real}} = 0.65$ ; $w_{\text{noisy}} = 0.35$; $\beta = w_{\text{KL}} \cdot w_{\text{as}}$, where $w_{\text{KL}} = 0.3$ and $w_{\text{as}} = 1/\text{\tt epoch\_num}$ if $\text{\tt epoch\_num} < as = 25$, else $1$.

\section{Experimental Results}

\subsection{Baseline}
The following classifiers are used as baseline models for comparison with our approach, selected to evaluate their performance both in terms of overall accuracy and their response to noisy instances in the dataset.

In the table \ref{tab:baseline_scores} are indicated the overall results. We choose the best one by accuracy to compare to our model that is Gradient Boosting.

\begin{table}[t]
\centering

\renewcommand{\arraystretch}{1.5} 

\begin{tabular}{lllll}
\toprule
Model                      & Accuracy & F1 & Precision & Recall \\
\midrule
AdaBoostClassifier         & 0.8191          & \textbf{0.8194} & \textbf{0.8765}    & 0.8191 \\
BaggingClassifier          & 0.8240          & 0.8106          & 0.8650             & 0.8240 \\
DecisionTreeClassifier     & 0.8191          & 0.8059          & 0.8555             & 0.8191 \\
GradientBoostingClassifier & \textbf{0.8264} & 0.8128          & 0.8664             & \textbf{0.8264} \\
KNeighborsClassifier       & 0.7726          & 0.7685          & 0.8380             & 0.7726 \\
LSTM                       & 0.5795          & 0.4252          & 0.3358             & 0.5795 \\
LogisticRegression         & 0.8240          & 0.8104          & 0.8621             & 0.8240 \\
MLP                        & 0.7946          & 0.7692          & 0.8166             & 0.7946 \\
RandomForestClassifier     & 0.8093          & 0.7966          & 0.8379             & 0.8093 \\
SVC                        & 0.8264          & 0.8128          & 0.8664             & 0.8264 \\
\bottomrule
\end{tabular}
\caption{Performance of baseline classifiers}
\label{tab:baseline_scores}
\end{table}

\begin{figure*}[b]
    \centering
    \includegraphics[width=\linewidth]{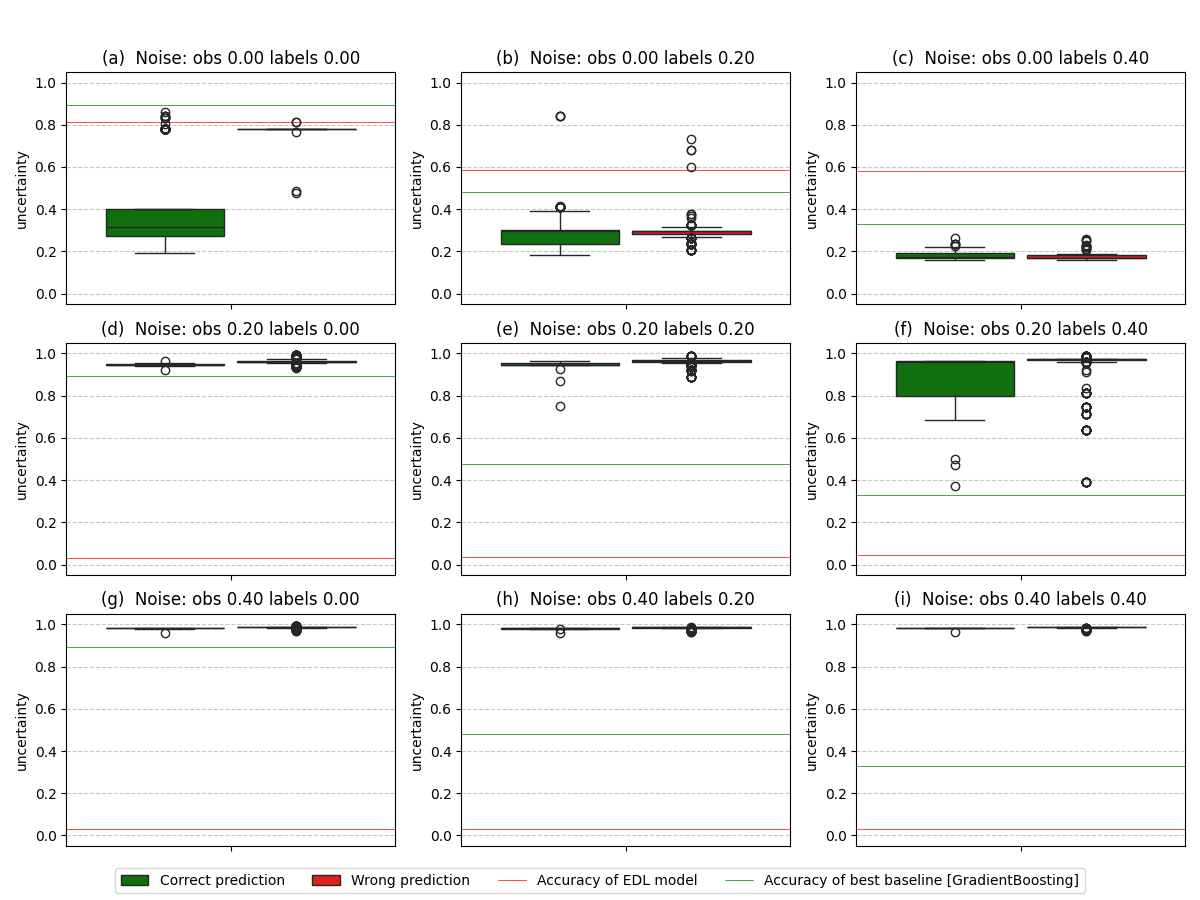}
    \caption{Uncertainty distribution as noise on observations and labels varies.}
    \label{fig:results_boxplot}
\end{figure*}

\begin{description}
    \item[AdaBoost] Adaptive boosting algorithm that combines multiple weak learners sequentially.
    Each subsequent learner focuses on correcting the errors made by previous learners.
    \item[Bagging] Bootstrap aggregating method that trains multiple models on different subsets of data.
    Reduces variance by averaging predictions from independent base learners.
    \item[Decision Tree] Interpretable tree-based model that makes predictions using branching rules.
    Creates hierarchical decision boundaries based on feature value thresholds.
    \item[Gradient Boosting] Sequential boosting technique that uses gradient descent optimization.
    Builds models iteratively to minimize residual errors from previous iterations.
    \item[K Nearest Neighbors] Instance-based lazy learning algorithm that stores all training data.
    Makes predictions by finding the k most similar instances in feature space.
    \item[Logistic Regression] Linear classifier that uses sigmoid activation for probability estimation.
    Applies linear transformation followed by logistic function for binary classification.
    \item[Random Forest] Ensemble method combining multiple decision trees with random feature selection.
    Reduces overfitting through bootstrap sampling and feature randomization.
    \item[SVC] Support Vector Classifier that finds optimal decision boundaries using kernel methods.
    Maximizes margin between classes while handling non-linear relationships through kernels.
    \item[MLP] Multi-layer perceptron feedforward neural network with hidden layers.
    Uses backpropagation to learn non-linear mappings between input and output.
    \item[LSTM] Long Short-Term Memory recurrent neural network for sequential data.
    Handles long-term dependencies through gating mechanisms and memory cells.
\end{description}

\subsection{Uncertainty-Aware Attack Stage Classification}

We now present the main experiment, designed to evaluate the model's capacity to detect attack stages under conditions of uncertainty. The training phase includes both clean data samples and out-of-distribution (OOD) examples. The latter are generated by applying 40\% bit-level noise to original instances. We chose 40\% bit-level noise because preliminary analysis showed that this level causes baseline models to achieve classification performance similar to random. This approach is intended to reflect realistic scenarios in which input data may be incomplete, corrupted, or otherwise unreliable.

During evaluation, the model is tested on noisy data constructed as follows: for a given noise level $X\%$, each bit in an input instance \--- comprising both observation and label bits \--- is independently flipped with probability $X$. This method ensures that noise is applied uniformly across all components of the input-output pair. Testing is conducted across three levels of noise: 0\%, 20\%, and 40\%.

In the absence of noise (Figure~\ref{fig:results_boxplot}a), the model demonstrates a clear distinction between uncertainty levels for correct and incorrect predictions. Correct classifications exhibit low uncertainty, while incorrect ones are associated with higher uncertainty. This separation suggests that uncertainty estimates can serve as a useful metric for filtering unreliable predictions, thereby improving decision-making robustness.

\begin{figure}
    \centering
    \includegraphics[width=0.88\linewidth]{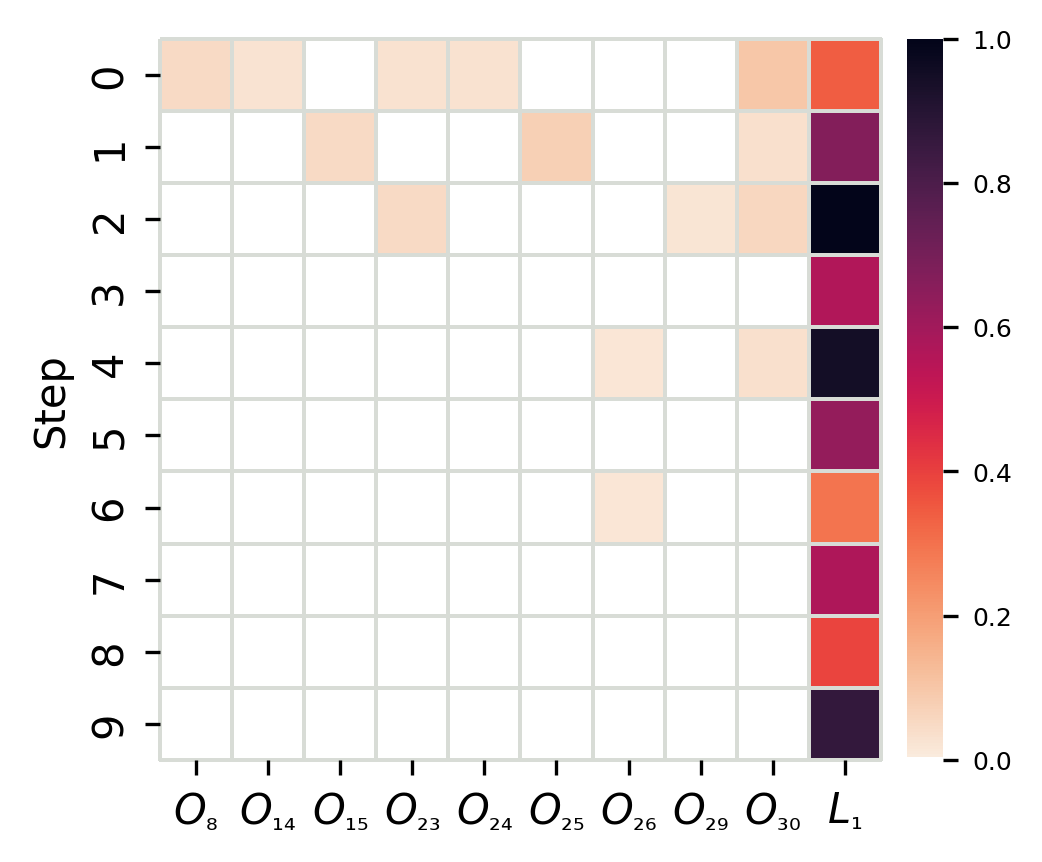}
    \caption{Heatmap of the mean relevance SHAP values for observations ($o_1$, \ldots, $o_{30}$) and labels ($l_1$ and $l_2$). Columns of features with value 0 for every step are omitted.}
    \label{fig:shap}
\end{figure}

As label noise increases (Figure~\ref{fig:results_boxplot}b and \ref{fig:results_boxplot}c), the uncertainty associated with correct and incorrect predictions begins to converge. This convergence reduces the discriminative utility of uncertainty values. The behaviour is partly attributable to dataset characteristics. In particular, the model assigns greater importance to the first label bit (feature $L_{1}$ in the SHAP feature map; see Figure~\ref{fig:shap}) and lower importance to the final bit $L_{2}$. This prioritisation results from class imbalance among the three possible RM states (Figure~\ref{ctf_rm}). Under noisy conditions, certain bit-flip combinations \--- such as all-zero labels or specific corrupted patterns \--- may still resemble common in-distribution cases. Consequently, the model fails to identify these samples as anomalous, and the associated uncertainty remains low.

The relatively small size of the label vector, in comparison to the observation vector, further biases the model towards weighting noise in the observations more heavily. This trend is evident in Figure~\ref{fig:results_boxplot}g, \ref{fig:results_boxplot}h and \ref{fig:results_boxplot}i, where increasing observation noise leads to a measurable rise in uncertainty. The model consistently identifies such instances as OOD, which indicates a degree of robustness against corrupted input features.

As noise levels increase, a general decline in classification accuracy is observed. In high-noise conditions, model performance falls below the expected random baseline of $1/n_{\text{class}}$. This result is attributed to a bias towards the minority class, which is introduced during training on an imbalanced dataset. Specifically, the model tends to misclassify instances in favour of the class with fewer training samples (the third RM state in this case). This effect is not present when the dataset is artificially balanced, although further experimentation was not pursued due to insufficient data volume.

Finally, it is noted that the best-performing baseline model exhibits a more pronounced decline in accuracy under increasing noise. This behaviour is likely a consequence of greater dependence on label information, which is more severely degraded by noise when compared to the EDL-based approach.

\section{Conclusions}
This work has explored the problem of attack stage inference in Advanced Persistent Threat scenarios, with a focus on incorporating uncertainty awareness through Evidential Deep Learning. By modelling predictive uncertainty via a Dirichlet distribution, the approach enables principled handling of out-of-distribution inputs without the need for explicit OOD data during training. Initial experiments in a simulated environment indicate that the proposed method can provide calibrated predictions and identify anomalous inputs, supporting more informed and adaptive response strategies.

Future work will involve a more comprehensive statistical evaluation of the proposed framework. This includes experiments using multiple random seeds to account for variance in model training, and the assessment of performance across a broader set of scenarios that more closely reflect real-world heterogeneity in attack behaviours and system environments. Additionally, a detailed analysis of the applicability and limitations of EDL in the context of operational cyber defence will be conducted, with emphasis on its integration into existing detection and response workflows. These steps are necessary to establish the reliability and generalisability of the method in practical deployments.



\begin{ack}
This project was partially funded by the Italian Ministry of University as part of the PRIN: PROGETTI DI RICERCA DI RILEVANTE INTERESSE NAZIONALE – Bando 2022, Prot. 2022EP2L7H
This work was partially supported by project SERICS (PE00000014) under the MUR National Recovery and Resilience Plan funded by the European Union – NextGenerationEU, specifically by the project NEACD: Neurosymbolic Enhanced Active Cyber Defence (CUP J33C22002810001).
This work was partially supported by the European Office of Aerospace Research \& Development (EOARD) under award number FA8655-22-1-7017 and by the US DEVCOM Army Research Laboratory (ARL) under Cooperative Agreements \#W911NF2220243 and \#W911NF1720196. Any opinions, findings, and conclusions or recommendations expressed in this material are those of the author(s) and do not necessarily reflect the views of the authors or of the United States government.
%
\end{ack}



\bibliography{biblio}

\end{document}